\documentclass[prb,aps,twocolumn,superscriptaddress]{revtex4}
\usepackage{amsmath}
\usepackage{color}
\usepackage{bbm}
\usepackage{amssymb}
\usepackage{epsfig}
\usepackage{multirow}
\usepackage{amsbsy}
\usepackage{array}
\usepackage{diagbox}
\usepackage{bm}
\usepackage{extarrows}
\usepackage{graphicx}
\usepackage{subfigure}
\usepackage{appendix}
\usepackage{txfonts}
\usepackage{lipsum}
\usepackage{bbding}
\usepackage{pifont}
\usepackage{makecell}
\usepackage[version=4]{mhchem}
\allowdisplaybreaks[4]
\usepackage[colorlinks=true,linkcolor=blue,citecolor=blue,urlcolor=blue,bookmarks=false]{hyperref}
\begin{document}

\title{Inversion-symmetric Electron Gases as New Platforms for Topological Planar Josephson Junctions}
 
\author{Jiong Mei}
\affiliation {Institute of Physics, Chinese Academy of Sciences, Beijing 100190, China}
\affiliation{University of Chinese Academy of Sciences, Beijing 100190, China}

\author{Kun Jiang}
\affiliation {Institute of Physics, Chinese Academy of Sciences, Beijing 100190, China}
\affiliation{University of Chinese Academy of Sciences, Beijing 100190, China}

\author{Shengshan Qin}\email{qinshengshan@bit.edu.cn}
\affiliation{School of Physics, Beijing Institute of Technology, Beijing 100081, China}

\author{Jiangping Hu}
\email{jphu@iphy.ac.cn}
\affiliation{Beijing National Research Center for Condensed Matter Physics, and Institute of Physics, Chinese Academy of Sciences, Beijing 100190, China}
\affiliation{Kavli Institute for Theoretical Sciences and CAS Center for Excellence in Topological Quantum Computation, University of Chinese Academy of Sciences, Beijing 100190, China}
\affiliation{New Cornerstone Science Laboratory, 
	Beijing, 100190, China}
	\begin{abstract}
Intrinsic Rashba spin-orbital coupling (SOC) can exist in  centrosymmetric materials with local inversion symmetry breaking. Here we show that  such a SOC can induce topological superconductivity together with  an in-plane Zeeman field in planar Josephson junctions formed by the centrosymmetric materials. A single Majorana mode can be created at each end of the junction. We demonstrate this result in a model  based on iron-based superconductors. We derive the necessary Fermi surface condition for the topological planar junction and calculate the topological phase diagram with respect to the in-plane Zeeman field and the phase difference between the two superconductors. We provide experimental characteristics for the topological superconductivity, including the differential conductance and the Fano factor tomography which can be measured in the scanning tunneling spectroscopy.  Our study reveals that the centrosymmetric systems with local-inversion-symmetry breaking can serve as new platforms for the topological planar Josephson junctions, and help to find more experimentally feasible materials for the topological superconductors.
\end{abstract}
	
\maketitle

\textit{Introduction.}—Majorana mode is a special kind of quasiparticle whose antiparticle is the particle itself, and it usually appears at the topological defects or on the edges of topological superconductors\cite{RevModPhys.83.1057,RevModPhys.88.035005,alicea2012new,kitaev2001unpaired,read2000paired,PhysRevB.82.115120}. The search for the Majorana modes is not only of fundamental scientific significance, but also of great importance for the realization of fault-tolerant quantum computing\cite{nayak2008non,doi:10.1146/annurev-conmatphys-030212-184337}. Over the past decades, tremendous efforts have been made in the study of the topological superconductors\cite{fu2008superconducting,PhysRevLett.104.040502,lutchyn2010majorana,PhysRevLett.105.177002,PhysRevLett.105.046803,PhysRevLett.107.097001,PhysRevLett.111.047006,PhysRevLett.111.056402,PhysRevB.88.155420,nadj2013proposal,PhysRevLett.115.127003,Li2016,xu2016topological,PhysRevB.93.224505,PhysRevX.9.011033,PhysRevLett.111.087002,doi:10.7566/JPSJ.82.113707,PhysRevLett.111.056403,PhysRevLett.112.106401,PhysRevLett.115.187001,PhysRevLett.123.027003,PhysRevLett.122.207001,QIN20191207,PhysRevB.102.094503,PhysRevB.93.020505,PhysRevLett.122.227001,PhysRevB.81.134508,doi:10.1126/sciadv.aaz8367,PhysRevLett.121.096803,PhysRevLett.121.186801,PhysRevLett.122.187001,PhysRevX.10.041014,10.1093/nsr/nwy142}, and great progresses have been achieved\cite{PhysRevLett.107.217001,lutchyn2018majorana,Das2012,doi:10.1126/science.1259327,doi:10.1126/science.1216466,PhysRevLett.112.217001,PhysRevLett.114.017001,doi:10.1126/science.aan4596,doi:10.1126/science.aao1797,PhysRevX.8.041056,Kong2019,Machida2019,Kong2021,doi:10.1126/science.aaw8419,doi:10.1126/science.aav3392,Li2022} especially in the artificial superconducting heterostructures\cite{lutchyn2018majorana,Das2012,doi:10.1126/science.1259327,doi:10.1126/science.1216466,PhysRevLett.112.217001,PhysRevLett.114.017001,doi:10.1126/science.aav3392}. One famous proposal for the topological superconductivity is the Fu-Kane proposal\cite{fu2008superconducting}. It is found that in proximity with conventional superconductors, the superconducting surface Dirac cone of a 3D strong topological insulator becomes a topological superconductor and a single Majorana zero-energy mode (MZM) is expected in each vortex on the surface. In experiments, the superconducting surface Dirac cone has been verified in $\ce{Bi_2Te_3}/\ce{NbSe_2}$ heterostructure\cite{doi:10.1126/science.1216466} and in some iron-based superconductors\cite{doi:10.1126/science.aan4596}, and evidence for the vortex bound MZMs has been detected\cite{PhysRevLett.112.217001,PhysRevLett.114.017001,doi:10.1126/science.aao1797,PhysRevX.8.041056,Kong2019,Machida2019,Kong2021,Li2022}. Many other superconducting heterostructures have also been proposed to realize the topological superconductivity and signatures for the MZMs have been observed experimentally, such as the 1D Rashba nanowires proximity to conventional superconductors\cite{lutchyn2010majorana,PhysRevLett.105.177002,Das2012}, magnetic atomic chain deposited on superconductors\cite{PhysRevB.88.155420,nadj2013proposal,PhysRevLett.115.127003,Li2016,doi:10.1126/science.1259327}, etc.

Josephson junctions are another promising artificial heterostructures to realize the topological superconductivity and the MZMs\cite{fu2008superconducting,PhysRevLett.115.237001,PhysRevLett.122.126402,PhysRevLett.122.107701}. Recently, the planar Josephson junction constructed by a 2D electron gas with strong spin-orbit coupling and a conventional superconductor, has been suggested to be a  platform for engineering topological superconductivity\cite{pientka2017topological,hell2017two}; with an external magnetic field, the planar junction can host topological superconductivity. Interestingly, in such a topological junction the MZMs can be conveniently controlled by the superconducting phase, and in the $\pi$-junction condition even a rather small magnetic field can drive the system into the topologically nontrivial state. Moreover, a setup based on the topological Josephson junctions can be convenient to realize topological superconductor networks implementing the non-Abelian braiding\cite{hell2017two,zhou2020phase}. In recent experiments, signatures for the existence of the MZMs in the planar Josephson junctions have been observed\cite{ren2019topological,fornieri2019evidence}.

However, for a long time the construction of the topological junctions has been limited to the noncentrosymmetric Rashba electron gas, which is  an obstruction in the pursuit of the MZMs. In this work, we generalize the material candidates for the topological planar Josephson junctions to the centrosymmetric systems. We focus on materials with local-inversion-symmetry-breaking crystal structures\cite{doi:10.1146/annurev-conmatphys-040521-042511} which are globally centrosymmetric but allows intrinsic Rashba spin-orbit coupling, and demonstrate that such centrosymmetric materials can serve as new platforms for the topological planar Josephson junctions.


\begin{figure}[!htbp]
		\centering
		\epsfig{figure=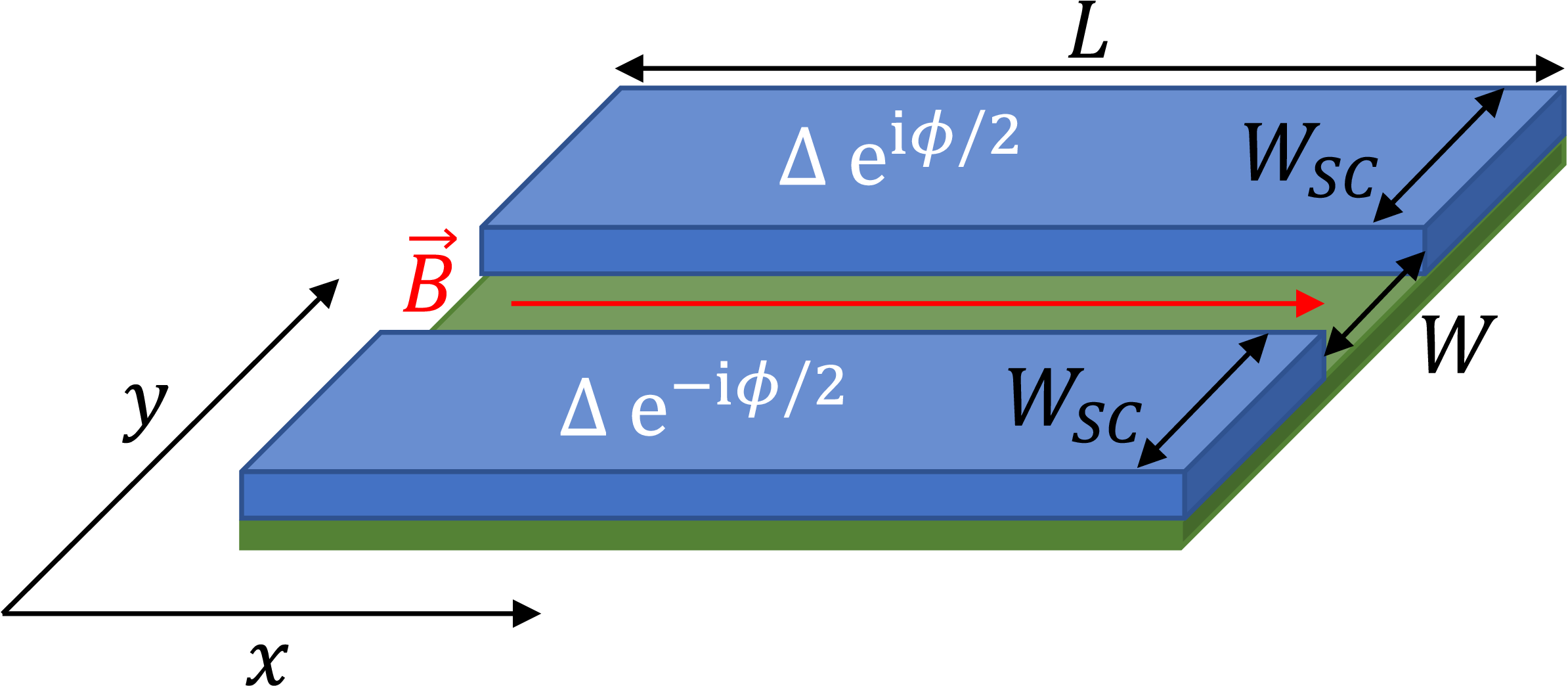, width=0.35\textwidth}
		\caption{The setup of a topological planar Josephson junction. In the work, we consider a 2D centrosymmetric electron gas (green) is in proximity with two separate conventional superconductors with a phase difference $\phi$. An in-plane magnetic field $\vec{B}$ is applied parallel to the junction.}
		\label{fig:setup}
\end{figure}


\textit{Model.}—We first use a simple model to demonstrate the above assertion. We consider the planar Josephson junction sketched in Fig.\ref{fig:setup}, which is composed with an electron gas and two superconductor leads. It has been shown that such Josephson junctions are promising to realize topological superconductivity, when the electron gas has strong Rashba spin-orbit coupling stemming from the absence of inversion symmetry\cite{pientka2017topological,hell2017two}. Different from the previous studies, in our consideration we focus on the inversion-symmetric electron gas. Specifically, we consider the electron gas in materials with local-inversion-symmetry-breaking crystal structures. One key feature for such materials is that though they are globally inversion symmetric, the materials can have strong Rashba spin-orbit coupling due to the intrinsic electric dipole arising from the special crystal structures\cite{doi:10.1146/annurev-conmatphys-040521-042511,Zhang2014,qin2022topological}. Therefore, we expect such centrosymmetric materials share some similarities with the noncentrosymmetric ones. In the specific model, we consider an electron gas respecting similar crystal structures with the iron-based superconductors, which is a typical structure with local-inversion-symmetry breaking, and the Hamiltonian reads as\cite{qin2022topological,PhysRevLett.119.267001}
\begin{align}\label{eqn:kspace H}
		H_0(\boldsymbol{k})=&2t(\cos k_{x} +\cos k_{y})\sigma_{0}s_{0}+2R\sin k_{x}\sigma_{3}s_{2} \notag \\
       &+2R\sin k_{y}\sigma_{3}s_{1}
		+4t_{1}\cos\frac{k_{x}}{2}\cos\frac{k_{y}}{2}\sigma_{1}s_{0},
\end{align}
where $s$, $\sigma$ are Pauli matrices representing the spin and sublattice degrees of freedom respectively. Actually, the Hamiltonian in Eq.\eqref{eqn:kspace H} describes a condition where a $s$ orbital is put on each iron lattice site in the monolayer \ce{FeSe}. In the model, $t$ is the nearest-neighborhood (NN) intrasublattice hopping, $t_{1}$ is the NN intersublattice hopping, $R$ is the NN intrasublattice Rashba spin-orbit coupling which maintains the inversion symmetry. The bands depicted by $H_0(\boldsymbol{k})$ are doubly degenerate due to the inversion symmetry and the time reversal symmetry existing simultaneously. It is worth mentioning that the symmetry group of the crystal enforces fourfold band degeneracy at the time reversal invariant points except $(0, 0)$ in the Brillouin zone, leading to that the Fermi surfaces around these points always come in pairs\cite{cvetkovic2013space,zhang2023symmetryprotected,qin2022spintriplet}. In the following, we choose the the parameters in Eq.\eqref{eqn:kspace H} as $t=-1$, $t_{1}=0.7$, $R=0.6$, and set the chemical potential $\mu=-2.2$ to make the electron gas merely have one twofold degenerate Fermi surface surrounding $(0, 0)$.
	
The two superconductor leads provide conventional superconductivity for the above electron gas through the proximity effect, and the superconductivity in the two leads have a phase difference $\phi$ which can be feasibly controlled by an external magnetic flux, as illustrated in Fig.\ref{fig:setup}. Accordingly, the superconductivity in the electron gas can be depicted by the following Hamiltonian
\begin{align}\label{eqn:junction sc}
	H_{SC} = &\Delta(\boldsymbol{r})\tau_+\sigma_0s_0 + \Delta^*(\boldsymbol{r})\tau_-\sigma_0s_0,
\end{align}
where $\tau$ is the Pauli matrix in the Nambu space with $\tau_{\pm}=(\tau_1\pm i\tau_2)/2$, and the Hamiltonian is written in the basis $(\psi_{\boldsymbol{k}}, is_2\psi_{-\boldsymbol{k}}^{\dagger})$ with $\psi_{\boldsymbol{k}}$ being the basis of $H_0(\boldsymbol{k})$ in Eq.\eqref{eqn:kspace H}. In Eq.\eqref{eqn:junction sc}, $\Delta(\boldsymbol{r}) = \Delta  e^{i\mathrm{sgn(y)\phi/2}} \Theta(|y|-W/2)$, where $\mathrm{sgn()}$ is the sign function, $\Theta()$ the step function, and $W$ the width of the junction as shown in Fig.\ref{fig:setup}.

We also consider an in-plane magnetic field applied parallel to the junction as illustrated in Fig.\ref{fig:setup}, leading to a Zeeman splitting in the electron gas which can be described as
\begin{align}\label{eqn:junction Zeeman}
	H_{Z} = & E_Z(y) \sigma_0 s_1.
\end{align}
In the above equation, the Zeeman energy takes the form $E_Z(y)=g(y)\mu_BB/2$, with $B$ being the strength of the magnetic field, $g(y)$ the Lande factor and $\mu_B$ the Bohr magneton. In general, the $g$ factors can be different for that in the junction and that underneath the superconductor leads. Therefore, the Zeeman energy of the electron gas takes the form $E_Z(y) = E_{Z,L}\theta(|y|-W/2) + E_{Z,J}\theta(W/2-|y|)$. In the following, for simplicity we assume a zero Zeeman energy for electron underneath the leads, and discuss the nonzero $E_{Z,L}$ case in the SM.

\textit{Topological superconductivity.}—In the planar Josephson junction in Fig.\ref{fig:setup}, Andreev reflection occurs at the interface between the normal region and the superconducting region, and bound states develop in the junction region, i.e. the intermediate region between the two superconductor leads. These Andreev bound states form a quasi-1D system along the junction, and we can consider the topological property of the system. Moreover, the in-plane magnetic field and the phase difference between the superconductor leads break the time reversal symmetry of the Josephson junction. Therefore, the quasi-1D system belongs to class $\mathrm{D}$ according to the Altland-Zirnbauer classification whose topological property is characterized by a $\mathbb{Z}_2$ topological index\cite{PhysRevB.78.195125,Ryu_2010}, and in the topologically nontrivial condition one single MZM exists at each end of the junction.

\begin{figure}[!htbp]
        \centering
        \epsfig{figure=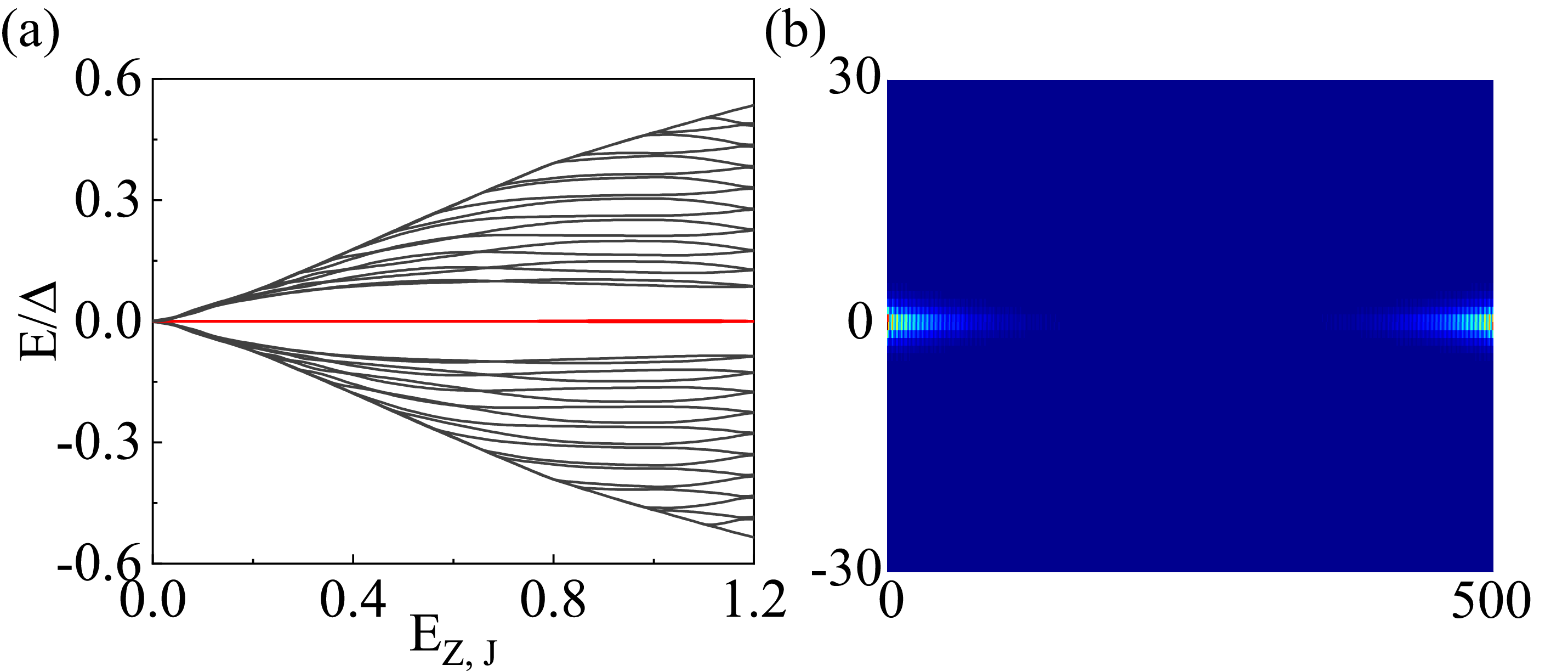, width=0.47\textwidth}
	\caption{(a) shows the quasiparticle energy spectrum for the planar Josephson junction in Fig.\ref{fig:setup} with open boundary conditions in both the $x$ and $y$ directions. The inset shows the Fermi surface of the electron gas in the normal state in the absence of magnetic field. (b) shows the wavefunction profiles for the zero energy modes in (a) at $E_Z=0.4$. The parameters are chosen as $t=-1$, $t_{1}=0.7$, $R=0.6$, $\phi=\pi$, $\Delta=0.3$, and the chemical potential is $\mu=-2.2$. In the calculation, we set $W=2$, $N_{x}=500$, $N_{y}=64$.}
	\label{fig2}
\end{figure}

To demonstrate the topological superconductivity in the Josephson junction, we solve the system straightforwardly in the finite condition, i.e. open boundary conditions in both the $x$ and $y$ directions in Fig.\ref{fig:setup}. By diagonalizing the whole BdG Hamiltonian numerically, we obtain the quasiparticle energy spectrum and find a pair of zero-energy modes with each located at each end of the junction, as shown in Fig.\ref{fig2}. Obviously, the zero-energy modes are the expected MZMs, which is a direct evidence for the topological superconductivity.


It is worth mentioning that in the calculations we find that the topological property of the junction strongly depends on the normal-state Fermi surface condition of the electron gas, and the topological superconductivity exist only when the number of Fermi surfaces is odd. We ignore the Kramers' degeneracy which is guaranteed by the inversion and time reversal symmetries, when we count the number of the Fermi surfaces, and more details are presented in the SM.

\begin{figure}[!htbp]
	\centering
        \epsfig{figure=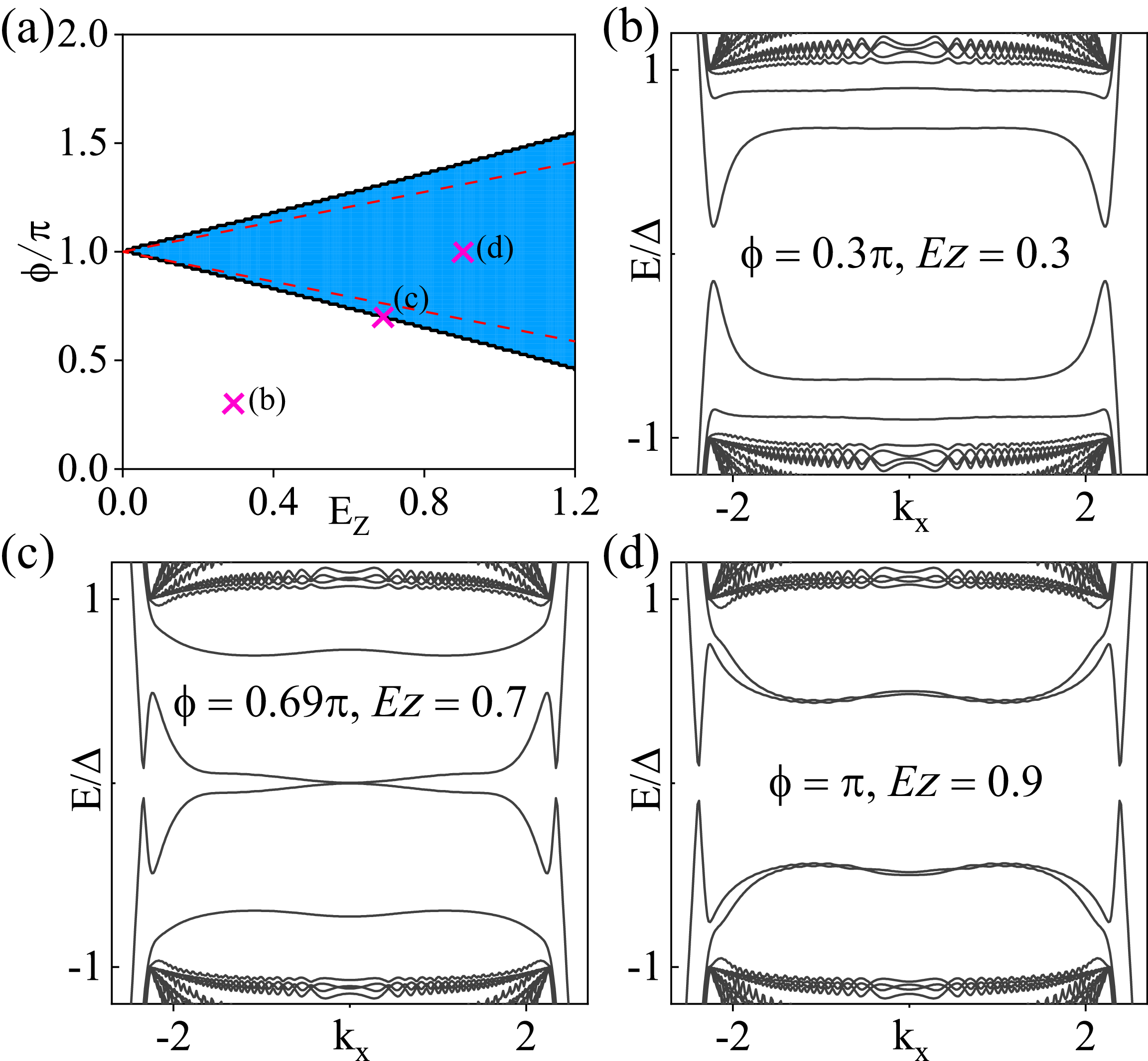, width=0.47\textwidth}
	\caption{(a) shows the topological phase diagram of the planar Josephson junction with respect to the phase difference $\phi$ and the magnetic field $\vec{B}$ with the blue (white) region representing the topologically nontrivial (trivial) region. The red dashed line is the phase boundary obtained by the analytical formula in Eq.\eqref{eqn:phase boundary}. (b)$\sim$(d) show the superconducting energy spectrum corresponding to the three points marked in the phase diagram in (a). In the calculations, the parameters are chosen as $t=-1$, $t_{1}=0.7$, $R=0.6$, $\mu=-2.2$, $\Delta=0.3$, $W=2$, $N_{y}=64$. }
	\label{fig3}
\end{figure}

\textit{Phase diagram.}—We calculate the phase diagram for the topological superconductivity in the planar Josephson junction with respect to the phase difference $\phi$ and the magnetic field $\vec{B}$. As mentioned, the Andreev bound states in the junction form a quasi-1D system which belongs to class $\mathrm{D}$ according to the Altland-Zirnbauer classification\cite{PhysRevB.78.195125,Ryu_2010}. The topological property of such a system is characterized by the $\mathbb{Z}_2$ topological index, which takes the formula $\mathbb{Z}_2=\mathrm{sgn} ( Pf [M(k_{x}=0)] Pf [M(k_{x}=\pi)] )$ with $Pf [M]$ standing for the pfaffian of the matrix $M$. In the expression, $M(k_{x})$ is the matrix of the BdG Hamiltonian of the whole junction in the Majorana representation which is antisymmetric due to the particle-hole symmetry existing at the time reversal invariant momentum. By calculating the $\mathbb{Z}_2$ topological index, we achieve the phase diagram for the topological superconductivity in Fig.\ref{fig3}(a). As shown, the topological superconducting state occupies a large region in the phase diagram, and in the $\pi$ junction condition a rather weak in-plane magnetic field can drive the junction into the topologically nontrivial phase. We also verify the phase diagram by simulating the superconducting energy spectrum  straightforwardly. As presented in Figs.\ref{fig3}(b)$\sim$(d), a gap-close-reopen process is clearly shown corresponding to the path in the parameter space in Fig.\ref{fig3}(a), indicating a topological phase transition at the phase boundary.


The topological superconductivity in the junction can also be analyzed analytically based on the scattering theory. Before carrying out the analysis, we first simplify the problem. According to the formula of the $\mathbb{Z}_2$ topological index, the topological property of the junction is thoroughly determined by the properties of the system at $k_x = 0$ and $k_x = \pi$. Moreover, considering the Fermi surfaces of the electron gas locate near $(0, 0)$, we can merely focus on the BdG Hamiltonian of the planar Josephson junction at $k_x = 0$. Namely, we study the gap-close-reopen process of the superconducting energy spectrum at $k_x = 0$. On the other hand, though the electron gas in Eq.\eqref{eqn:kspace H} is a two-band system (ignoring the Kramers' degeneracy), only one band crosses the Fermi energy with the other band far away from the Fermi energy. Therefore, it is reasonable to discard scattering related to the high-energy band and only consider the scattering from the low-energy band. With the above two points taken into consideration, the problem becomes a case similar to the 1D single-band superconductor-metal-superconductor junction, which is greatly simplified.

Such a problem can be conveniently solved in the continuum limit. At the interface between the normal region and the superconducting region, i.e. at $y = \pm W/2$ in Fig.\ref{fig:setup}, only pure Andreev reflection occurs with the scattering matrix being
\begin{equation}\label{eqn:S matirx@NS interface}
	s_{A,j}(\pm W/2)=e^{-i\beta}\begin{pmatrix}0 & e^{\pm i\phi/2}\\
		e^{\mp i\phi/2} & 0
	\end{pmatrix}.
\end{equation}
The scattering matrix is written in the basis $\begin{pmatrix}\psi_{j}^{e}(\pm k_{e}), & \psi_{j}^{h}(\mp k_{h})\end{pmatrix}^{T}$, where $e$ ($h$) stands for the electron (hole) and $j = 1, 2$ for the states from the two degenerate Fermi surfaces. In Eq.\eqref{eqn:S matirx@NS interface}, $\beta = \arccos (\frac{E}{\Delta})$ with $E$ being the energy of the Andreev bound states.
By solving the system in different regions and the boundary conditions at the interfaces, we can get the following equation
\begin{equation}\label{eqn:Andreev bound}
	e^{ik_{e,j}W}e^{-i\beta}e^{-i\phi/2}e^{ik_{h,j}(-W)}e^{-i\beta}e^{-i\phi/2}=1,
\end{equation}
After some algebra, we arrive at a simple equation satisfied by the Andreev bound states
\begin{equation}\label{eqn:subgap spectrum}
	\arccos(\frac{E}{\Delta})= -\frac{\phi}{2}+ \frac{\pi}{2}\frac{E \pm E_{Z}}{E_T} + n\pi,
\end{equation}
where $n$ is an integer, and $E_T = (\pi/2)v_F/W$ is the corresponding Thouless energy with $v_{F}=-2tk_F-\frac{2R^2k_F}{\sqrt{R^2k_F^2+4t_1^2}}$ and $k_F$ being the Fermi momentum. Based on Eq.\eqref{eqn:subgap spectrum}, we can immediately determine the boundary between the different topological phases by setting $E=0$, which turns out to be
\begin{equation} \label{eqn:phase boundary}
	\phi\,\text{mod}\,2\pi=(1\pm E_Z/E_T)\pi\,\text{mod}\,2\pi.
\end{equation}
We plot the phase boundary in Fig.\ref{fig3}(a) based on the analytical results in Eq.\eqref{eqn:phase boundary}, which matches with the numerical results well. More details on the analytical analysis are presented in the SM.

In the above analysis, we assume pure Andreev reflection at the interface between the normal region and the superconducting region. In fact, the topological phase diagram maintains qualitatively unchanged with the phase boundary slightly deformed, even if the normal reflection is considered at the interface. Compared with the previous study\cite{pientka2017topological}, the topological phase diagram for the junction constructed with centrosymmetric electron gas here is similar to that constructed with the Rashba electron gas, indicating some similarities between systems with local-inversion-symmetry breaking and the noncentrosymmetric systems.

Here, it is worth pointing out that the scattering theory provides an intuitive understanding on the correspondence between the topological superconductivity and the number of the Fermi surfaces of the electron gas. In the weak pairing condition, the scattering within the same Fermi surface dominates other processes, and we can approximately deal with the Andreev bound states from different Fermi surfaces separately. Since each Fermi surface contributes to one MZM at the end of the junction as shown in the above analysis, MZMs from different Fermi surfaces hybridize with each other, and only in the condition with odd-number Fermi surfaces the topological superconductivity in the junction can survive considering its $\mathbb{Z}_2$ classification.



\begin{figure}[!htbp]
        \centering
        \epsfig{figure=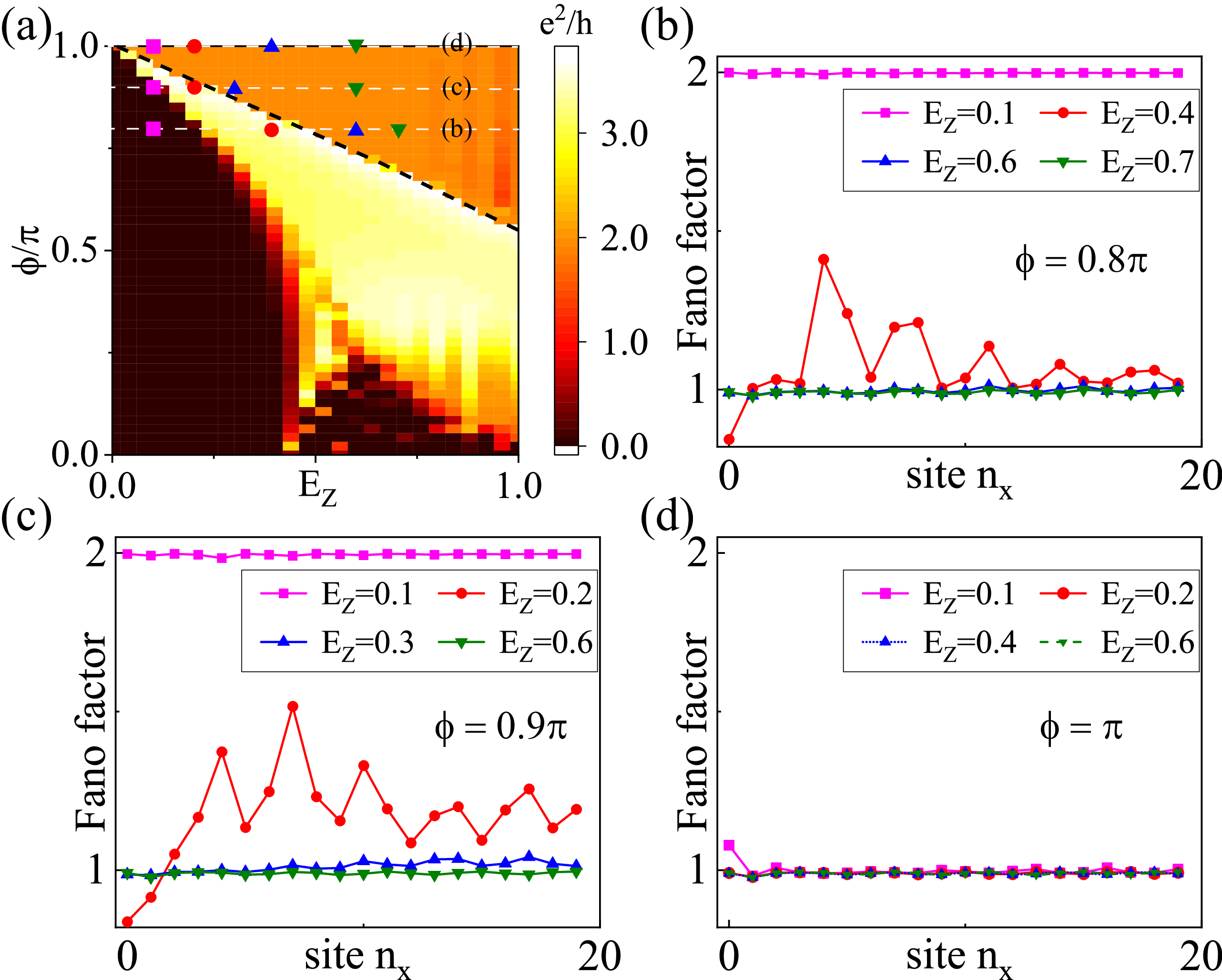, width=0.47\textwidth}
	\caption{ (a) shows the zero-bias conductance measured at the end of the junction, where the black dashed line indicates the topological phase boundary. Notice that in the region near the phase transition boundary, the zero-bias conductance exceeds $2e^2/h$ due to the contributions from the nearly gapless excitations in the junction as indicated in Fig.\ref{fig3}(c). (b)$\sim$(d) show the spatially resolved Fano factors as the tip moves along the $x$ direction with $n_x = 0$ located at the end of the junction, with respect to different $\phi$ and $E_Z$.
    In the conductance calculations, the energy width is set to $\Gamma=1.13$ and the temperature is $k_BT=\Delta/8000$, while in the simulations of the Fano factor tomography, we choose $\Gamma=0.3\Delta$ with a fixed bias voltage $eV_{\text{bias}}=0.01\Delta$ at the same temperature. The size of the system is set to be $N_x = 800$, $N_y = 64$. The other parameters of the junction Hamiltonian are the same as those in Fig. \ref{fig3}.  }
	\label{fig6}
\end{figure}

\textit{Experimental characterization.}—Finally, we present experimental evidences in scanning tunneling spectroscopy, including the measurements of the zero-bias conductance and the Fano factor tomography\cite{PhysRevB.104.L121406,mei2023identifying}. In the measurements, the coupling between the tip and the junction takes a general form
\begin{equation}\label{tunnelingH}
	H_{\text{tunnel}}=\sum_{i,s}[\tilde{t}_{i}c_{T,s}^{\dagger} c_{J,i,s} + H.c.],
\end{equation}
where $c_{T,s}$ ($c_{J,i,s}$) is the annihilation operator for the spin $s$ electron in the tip (junction) with $i$ standing for lattice site in the junction. Thus the tunneling events are characterized by the energy width $\Gamma = 2\pi\nu_T\sum_i\tilde{t}_{i}^2$, where $\nu_T$ is the density of states in the tip. The bias voltage $V$ between the tip and the junction is taken into account as $\mu_T = \mu + eV$, with $\mu_T$ ($\mu$) the chemical potential for the tip (junction). For simplicity, we assume the tunneling takes place merely between the tip and the lattice sites nearest to the tip in the junction in the simulations. Fig. \ref{fig6}(a) shows the zero-bias conductance in the scanning tunneling spectroscopy measurements. Apparently, The quantized zero-bias conductance $2e^2/h$ is a clear signal of the MZMs.


The Fano factor tomography is another useful method to distinguish the MBS from the trivial bound states. The spatially resolved Fano factors are defined as
\begin{equation}
	F(i,eV)=\frac{S(i,eV)}{2e\left|I(i,eV)\right|},
\end{equation}
where $I(i,eV)$ is the d.c.$\!$ tunneling current when the tip is located above lattice site $i$ in the junction, and $S(i,eV)$ is the corresponding shot noise. The shot noise is the zero-frequency limit of the time-symmetrized current-current correlator, $S = \int \mathrm{d}(t_1-t_2)S(t_1, t_2)$, where $S(t_1, t_2) = \left\langle \delta I(t_1)\delta I(t_2)\right\rangle + \left\langle \delta I(t_2)\delta I(t_1)\right\rangle$ and $\delta I(t) = I(t) - \left\langle I(t)\right\rangle $. In the high voltage regime $eV\gg\Gamma_i$, with $\Gamma_i = \Gamma\sum_{\sigma}\left(\left|u_{\sigma}(i)\right|^{2}+\left|v_{\sigma}(i)\right|^{2}\right)$ being the energy width $\Gamma$ multiplied with the probability of the low lying states at site $i$, the spatially resolved Fano factors represent the local particle-hole asymmetry of the bound states wave functions\cite{PhysRevB.104.L121406}.
For an isolated MZM which has perfect local particle-hole symmetry, the Fano factor takes a quantized value 1, while for a trivial bound state the spatially resolved Fano factors oscillate between 1 and 2. We calculate the Fano factors at different points in the parameter space ($\phi$, $E_Z$) and present the results in Figs.\ref{fig6}(b)$\sim$(d). As shown, in the topological region a flat plateau $F(n_x) = 1$ is observed in the vicinity of the end of the junction, signaling the existence of the MZM.

\textit{Discussion and conclusion.}—Recently, the centrosymmetric materials with local-inversion symmetry breaking have been systematically classified, and a large number of such kind of materials have heen identified\cite{Zhang2014,Wu2017,Zhang_2020}. As concrete examples, the iron-based superconductors are potential candidates for the construction of the topological planar Josephson junctions. $\ce{KFe_2As_2}$ and many 122 family of iron pnictides have three Fermi surfaces at the Brillouin zone center and two Fermi surfaces at the Brillouin zone corner\cite{Wu2024,Derondeau2017,PhysRevB.89.081103}, which satisfies the requirement of the topological superconductivity in the planar junctions. Besides the iron-based superconductors, the bilayer systems studied in Ref.\cite{nakosai2012topological} can also serve as a suitable platform to implement our proposal.

In summary, we show the centrosymmetric materials with local-inversion-symmetry-breaking crystal structures can serve as new material platforms to construct the topological planar Josephson junctions. This result  significantly extends   material candidates for  topological superconductors. Experimental evidences including the differential conductance and the Fano factor tomography in the scanning tunneling spectroscopy measurements, are provided to verify the MZMs bound in the junction. Our study can help to find more experimentally feasible materials to realize the long-pursuit MZMs.


\textit{Acknowledgment.}—This work is supported by the Ministry of Science and Technology (Grant No. 2022YFA1403900), the National Natural Science Foundation of China (Grant No. NSFC-12304163, No. NSFC-11888101, No. NSFC-12174428, No. NSFC-11920101005), the Strategic Priority Research Program of the Chinese Academy of Sciences (Grant No. XDB28000000, XDB33000000), the New Cornerstone Investigator Program, and the Chinese Academy of Sciences Project for Young Scientists in Basic Research (2022YSBR-048).

\nocite{*}
\bibliography{references}

\clearpage
\onecolumngrid
\begin{center}
	\textbf{\large Supplemental Materials}
\end{center}
\setcounter{equation}{0}
\setcounter{figure}{0}
\setcounter{table}{0}
\setcounter{page}{1}
\makeatletter
\renewcommand{\theequation}{S\arabic{equation}}
\renewcommand{\thefigure}{S\arabic{figure}}
\renewcommand{\bibnumfmt}[1]{[S#1]}
\renewcommand{\citenumfont}[1]{S#1}

\section{Tight-binding model and corresponding numerical results}
In this section, we present the tight-binding model and the corresponding numerical results for the single orbital model in the main text. The full Hamiltonian in real space can be represented as,

\begin{equation}
	H_{TB}=H_{normal}+H_{Z}+H_{\Delta}, \label{eqn:junction Hamiltonian}
\end{equation}
\begin{align}
	H_{normal}	=&-\mu\sum_{x,y}\left|x,y\right\rangle \left\langle x,y\right|\tau_{3}\sigma_{0}s_{0}+t\sum_{x,y}\left[\left|x+1,y\right\rangle \left\langle x,y\right|+\left|x,y+1\right\rangle \left\langle x,y\right|+H.C.\right]\tau_{3}\sigma_{0}s_{0}\notag \\
	&+	\left[iR\sum_{x,y}\left|x,y+1\right\rangle \left\langle x,y\right|\tau_{3}\sigma_{3}s_{1}+H.C.\right]+\left[iR\sum_{x,y}\left|x+1,y\right\rangle \left\langle x,y\right|\tau_{3}\sigma_{3}s_{2}+H.C.\right]\notag \\
	&+	t_{1}\sum_{x,y}\left|x,y\right\rangle \left\langle x,y\right|\tau_{3}\sigma_{1}s_{0}+t_{1}\sum_{x,y}\left[\left|x,y+1\right\rangle \left\langle x,y\right|+\left|x-1,y\right\rangle \left\langle x,y\right|+\left|x-1,y+1\right\rangle \left\langle x,y\right|\right]\tau_{3}m_{1}s_{0}\notag \\
	&+	t_{1}\sum_{x,y}\left[\left|x+1,y\right\rangle \left\langle x,y\right|+\left|x,y-1\right\rangle \left\langle x,y\right|+\left|x+1,y-1\right\rangle \left\langle x,y\right|\right]\tau_{3}m_{2}s_{0}
\end{align}
\begin{equation}
	H_{Z}=E_{Z}\sum_{y=(N_{y}-W)/2+1}^{y=(N_{y}+W)/2}\sum_{x}\left|x,y\right\rangle \left\langle x,y\right|\tau_{0}\sigma_{0}s_{1}
\end{equation}
\begin{equation}
	H_{\Delta}=\Delta e^{i\phi/2}\sum_{y=1}^{y=(N_{y}-W)/2}\sum_{x}\left|x,y\right\rangle \left\langle x,y\right|\left[\tau_{+}\sigma_{0}s_{0}+\tau_{-}\sigma_{0}s_{0}\right]+\Delta e^{-i\phi/2}\sum_{y=(N_{y}+W)/2+1}^{y=N_{y}}\sum_{x}\left|x,y\right\rangle \left\langle x,y\right|\left[\tau_{+}\sigma_{0}s_{0}+\tau_{-}\sigma_{0}s_{0}\right],
\end{equation}
where $m_{1}=(\sigma_{1}-i\sigma_{2})/2$, $m_{2}=(\sigma_{1}+i\sigma_{2})/2$. Here, we consider the simplest onsite s-wave pairing scenario.

\begin{figure}[!htbp]
	\begin{minipage}[b]{0.8\textwidth}
		\centering
		\subfigure[]{
			\epsfig{figure=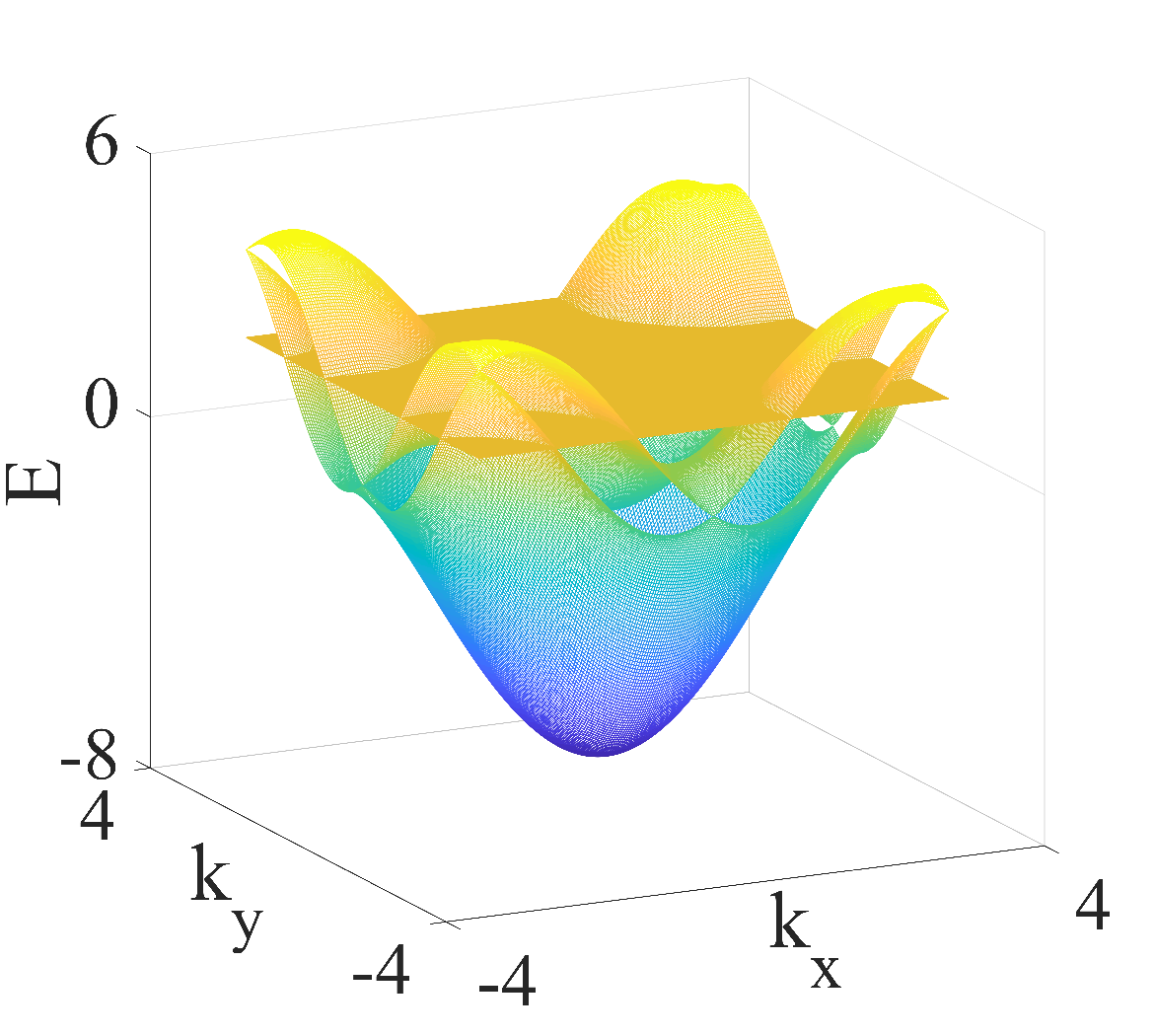, width=0.27\textwidth}
		}
		\subfigure[]{
			\epsfig{figure=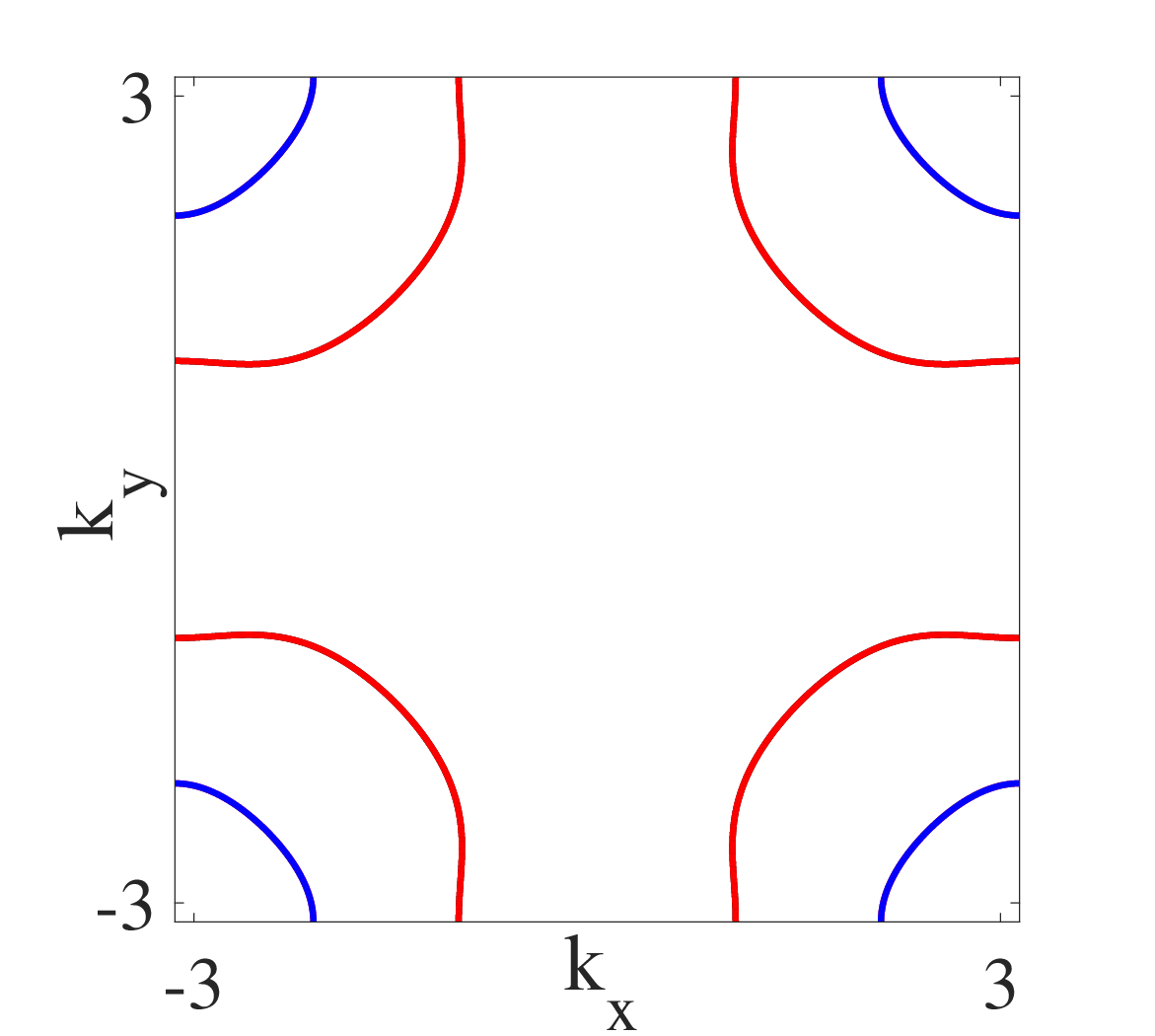, width=0.27\textwidth}
		}
            \subfigure[]{
			\epsfig{figure=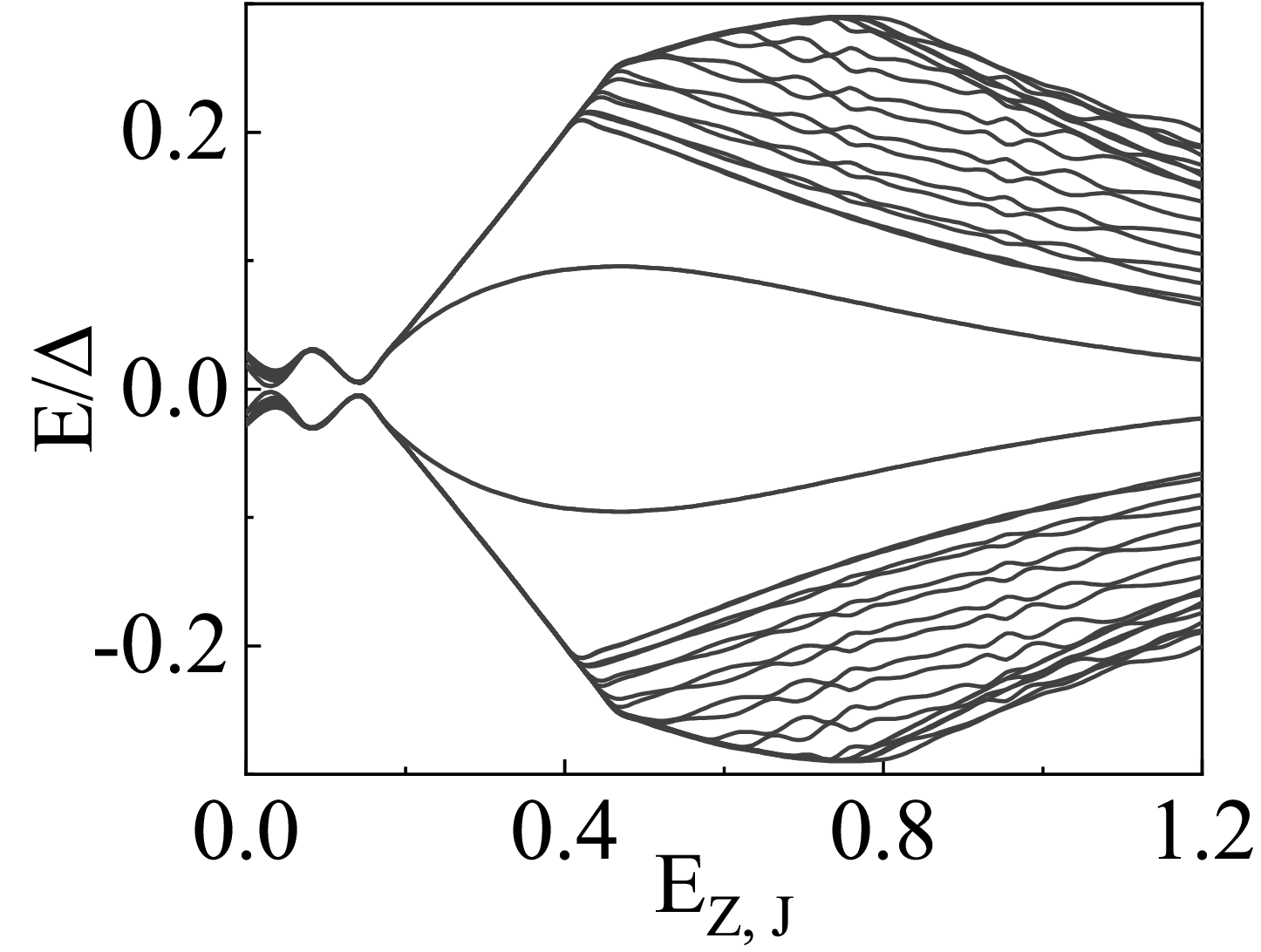, width=0.3\textwidth}
		}
	\end{minipage}
 
	\begin{minipage}[b]{0.8\textwidth}
		\centering
		\subfigure[]{
			\epsfig{figure=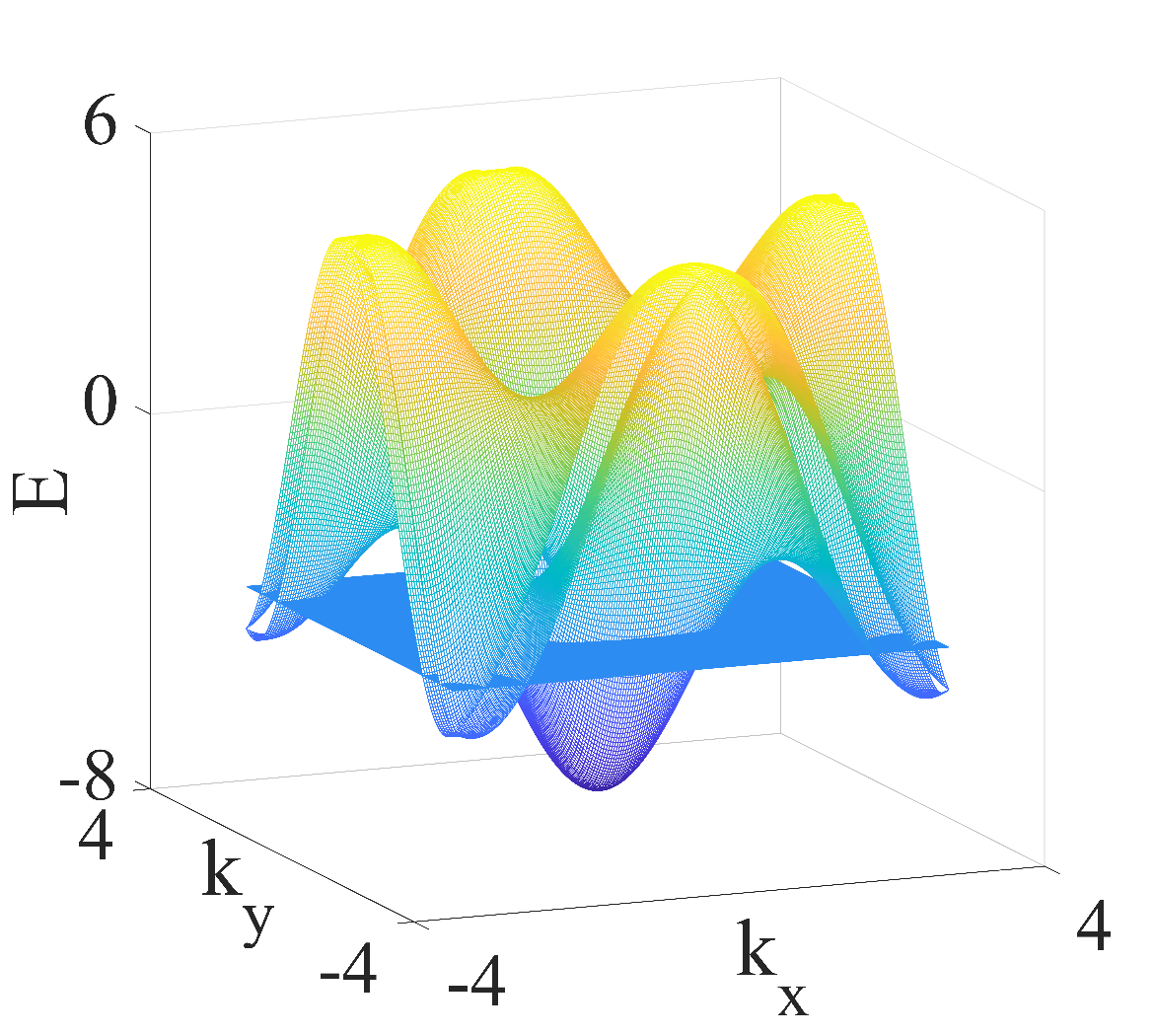, width=0.27\textwidth}
		}
		\subfigure[]{
			\epsfig{figure=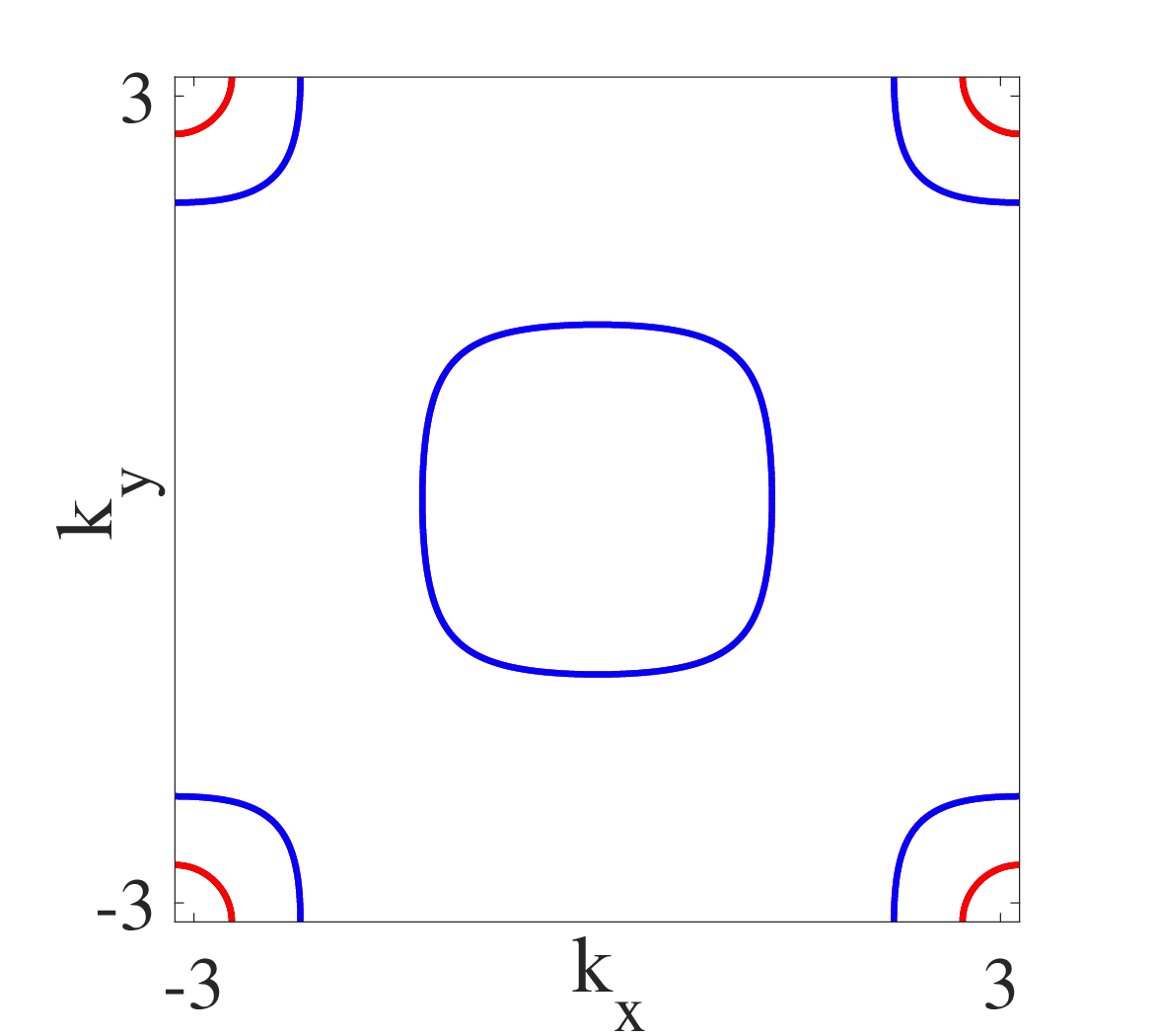, width=0.27\textwidth}
		}
            \subfigure[]{
			\epsfig{figure=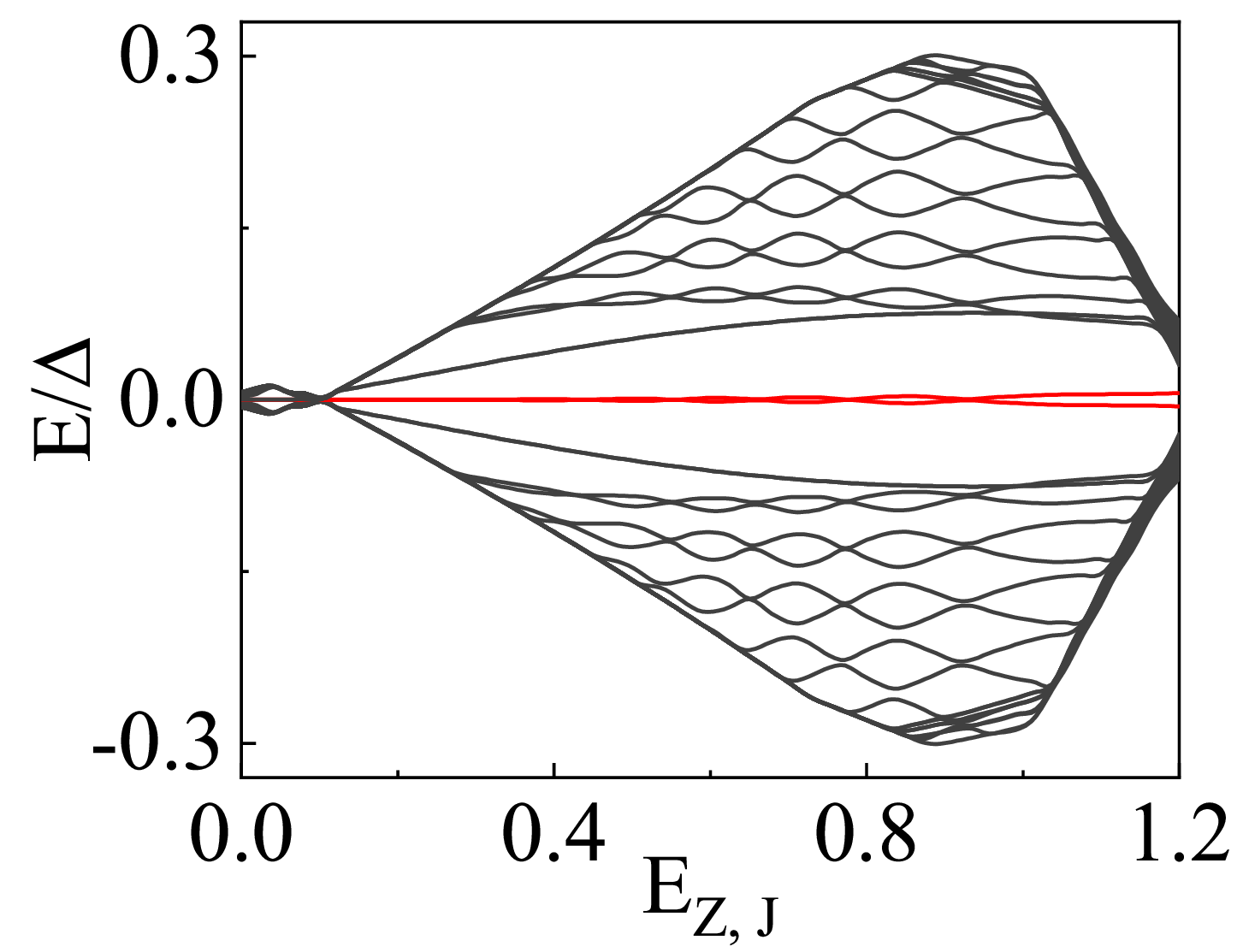, width=0.3\textwidth}
		}
	\end{minipage}
	\caption{(a)-(c) Double FSs around $M$ point. [(a) and (b)] The band structure and the corresponding FSs with parameters chosen as $\{t,\,t_1,\,R,\,\mu\}=\{-1,\,0.7,\,0.6,\,2\}$. (c) The spectrum of a $\pi\text{-}0$ junction under open boundary condition in the $x$ direction. The parameters for this junction are chosen as $\{\Delta,\,W,\,N_x,\,N_y\}=\{0.3,\,2,\,460,\,60\}$. (d-e) Three FSs with two around $M$ point and one around $\Gamma$ point. [(d) and (e)] The band structure and the corresponding FSs with parameters chosen as $\{t,\,t_1,\,t_2,\,R,\,\mu\}=\{-0.1,\,0.5,\,-1.2,\,0.6,\,-3.5\}$. Here $t_2$ represents the next NN intrasublattice hopping strength. (e) The open boundary spectrum of the $\pi\text{-}0$ junction in this case where the parameters are chosen as $\{\Delta,\,W,\,N_x,\,N_y\}=\{0.3,\,2,\,800,\,84\}$}.
	\label{fig:M2FSs}
\end{figure}

For two doubly degenerate FSs around $M$ point, as shown in Figs. \ref{fig:M2FSs}(a) and (b), we tried a specific set of parameters to show the possible Majorana bound states (MBSs) in the case of $\pi\text{-}0$ junctions [Fig. \ref{fig:M2FSs}(c)]. As  mentioned in the text, the two MBSs on the same boundary are coupled, leading to a shift to finite energies.

In Figs. \ref{fig:M2FSs}(d)-(e), we consider a case in which there are two FSs around the $M$ point and one FS around the $\Gamma$ point. Our numerical results indicate that odd FSs (in the absence of Zeeman field) can be utilized to host MZMs. We emphasize that this conclusion holds for other nodeless s-wave pairing scenarios, such as AB intersublattice pairing, and is not limited to the onsite s-wave pairing scenario considered throughout this paper.

		

\section{Analytical analysis based on scattering theory}
In this section, we perform an analytical analysis of the topological phase transition of the junction using scattering theory. Our focus is on the case of a single FS near the $\Gamma$ point, for which we utilize the continuum version of the Hamiltonian, as described by Eq.~\eqref{eqn:S5} below,
\begin{equation}\label{eqn:S5}
h_0(\bm{k})=t(k_{x}^{2}+k_{y}^{2})\sigma_{0}s_{0}+t_{1}\sigma_{1}s_{0}+\lambda k_{x}\sigma_{3}s_{2}+\lambda k_{y}\sigma_{3}s_{1} -\mu\sigma_0s_0.
\end{equation}
As depited in Fig.~\ref{fig:setup}, we consider a planar Josephson junction with in-plane magnetic field $E_Z$ along the $x$ direction. When working in the basis of $\left(c_{\bm{k}},\,is_{2}c_{-\bm{k}}^{\dagger}\right)^{T}$, the BdG Hamiltonian in the SC region is:
\begin{equation}
H_{SC}=\begin{pmatrix}h_{0} & \Delta\\
\Delta^{*} & -h_{0}
\end{pmatrix},
\end{equation}
and the BdG Hamiltonian in the normal region is:
\begin{equation}
H_{N}=\begin{pmatrix}h_{0}+E_{Z}\sigma_{0}s_{1} & 0\\
0 & -h_{0}+E_{Z}\sigma_{0}s_{1}
\end{pmatrix}.
\end{equation}
We set the chemical potential $\mu\in(-t_1,t_1)$ to keep only one FS at $\Gamma$ point. We denote the Fermi velocity at chemical potential $\mu$ is $v_F$ in the absence of the magnetic field and we assume it is basically unchanged by the applying magnetic field. At $k_x=0$, $h_0(k_y)=(tk_y^2-\mu)\sigma_0s_0+t_1\sigma_1s_0+\lambda k_y\sigma_3s_1$. We apply a gauge transformation in the spin space such that $s_{1}\rightarrow Us_{1}U^{\dagger}=s_{3}$. In this case,
\begin{equation}
h_{0}(k_{y})=(tk_{y}^{2}-\mu)\sigma_{0}s_{0}+\begin{pmatrix}\lambda k_{y} &  & t_{1}\\
 & -\lambda k_{y} &  & t_{1}\\
t_{1} &  & -\lambda k_{y}\\
 & t_{1} &  & \lambda k_{y}
\end{pmatrix}.
\end{equation}
The eigenvectors are
\begin{align}
E& = (tk_{y}^{2}-\mu)+\sqrt{(\lambda k_{y})^{2}+t_{1}^{2}}:\quad\begin{pmatrix}\cos\frac{\theta}{2}\\
0\\
\sin\frac{\theta}{2}\\
0
\end{pmatrix} \quad\text{and}\quad \begin{pmatrix}0\\
\sin\frac{\theta}{2}\\
0\\
\cos\frac{\theta}{2}
\end{pmatrix},\\
E & =(tk_{y}^{2}-\mu)-\sqrt{(\lambda k_{y})^{2}+t_{1}^{2}}:\quad v_{1}=\begin{pmatrix}-\sin\frac{\theta}{2}\\
0\\
\cos\frac{\theta}{2}\\
0
\end{pmatrix}\;\text{and}\;v_{2}=\begin{pmatrix}0\\
-\cos\frac{\theta}{2}\\
0\\
\sin\frac{\theta}{2}
\end{pmatrix},\label{eqn:S10}
\end{align}
where $\theta=\arctan(\frac{t_1}{\lambda k_y})$. We focus on the lower branch $\left|v_{1}\right\rangle $ and $\left|v_{2}\right\rangle$ (i.e., eigenstates in Eq.~\eqref{eqn:S10}). The Zeeman term restricted in the subspace under this gauge is $E_{Z}\sigma_0s_3=E_{Z}\left(\left|v_{1}\right\rangle \left\langle v_{1}\right|-\left|v_{2}\right\rangle \left\langle v_{2}\right|\right)$.

In the case of $k_x=0$, the scattering states in the normal region can be written as:
\begin{equation}\label{eqn:S11}
\psi_{N,\alpha}^{e,j}=\frac{1}{\sqrt{N_{ej,\alpha}}}\begin{pmatrix}v_{j\alpha}\\
0
\end{pmatrix}e^{ik_{ej,\alpha}y},\quad\psi_{N,\alpha}^{h,j}=\frac{1}{\sqrt{N_{hj,\alpha}}}\begin{pmatrix}0\\
v_{j\alpha}
\end{pmatrix}e^{ik_{hj,\alpha}y},
\end{equation}
where $\alpha=+/-$ denotes the direction of the wave vector (along $\pm y$), and $j=1,2$ denotes different modes near the Fermi surface. The eigenstates depend on the direction of $k_y$, as $\theta=\arctan(\frac{t_1}{\lambda k_y})$. Here, $\psi_{N,+}^{e,j}$, $\psi_{N,-}^{h,j}$ are up movers, while $\psi_{N,-}^{e,j}$, $\psi_{N,+}^{h,j}$ are down movers. The normalization factors $N_{e(h),j\alpha}$ are defined to ensure that the scattering matrices are unitary. Within the Andreev approximation, the quasiparticle current of an incident wave can be expressed as:
\begin{equation}\label{eqn:S12}
J_{P,\alpha}^{y}=v_{F,\alpha}\psi^{\dagger}\tau_{z}\psi.
\end{equation}
For solving Andreev bound states, one can use the quasiparticle current as the normalization factor (i.e., $N=|J_P^y|$ in Eq.~\eqref{eqn:S11}) of the incoming and outgoing states.

In the presence of magnetic field, the correction to the wave vectors are:
\begin{align}
k_{e1,\alpha}\approx k_{F,\alpha}+\delta k_{e1,\alpha},\quad\delta k_{e1,\alpha} & =\frac{E-E_{Z}}{\hbar v_{F,\alpha}},\label{eqn:S13}\\
k_{e2,\alpha}\approx k_{F,\alpha}+\delta k_{e2,\alpha},\quad\delta k_{e2,\alpha} & =\frac{E+E_{Z}}{\hbar v_{F,\alpha}}.\label{eqn:S14}
\end{align}
Note that under our Nambu basis, the hole part Hamiltonian at $\bm{k}$ is $-h_0(\bm{k})+E_{Z}\sigma_0s_3$. Thus, we have
\begin{align}
k_{h1,\alpha}\approx k_{F,\alpha}+\delta k_{h1,\alpha},\quad\delta k_{h1,\alpha} & =\frac{E-E_{Z}}{-\hbar v_{F,\alpha}},\label{eqn:S15}\\
k_{h2,\alpha}\approx k_{F,\alpha}+\delta k_{h2,\alpha},\quad\delta k_{h2,\alpha} & =\frac{E+E_{Z}}{-\hbar v_{F,\alpha}}.\label{eqn:S16}
\end{align}
On the other hand, the scattering states in the SC region are:
\begin{align}
\psi_{S,\alpha}^{up,j} & =\begin{pmatrix}v_{j\alpha}e^{i\alpha\beta}\\
v_{j\alpha}e^{-i\phi/2}
\end{pmatrix}e^{ik_{F,\alpha}y+i\kappa_{\alpha}(y-W)},\qquad y\geq W,\\
\psi_{S,\alpha}^{down,j} & =\begin{pmatrix}v_{j\alpha}e^{-i\alpha\beta}\\
v_{j\alpha}e^{i\phi/2}
\end{pmatrix}e^{ik_{F,\alpha}y-i\kappa_{\alpha}y},\qquad y\leq0.
\end{align}
The parameters $\kappa_\alpha$, $\beta$ are given by
\begin{align}
\kappa_{\alpha} & =\frac{\sqrt{E^{2}-\Delta^{2}}}{\alpha v_{F,\alpha}}\\
\beta & =\begin{cases}
\arccos\frac{E}{\Delta} & \text{if}\quad E<\Delta\\
-i\text{acosh}\frac{E}{\Delta_{0}} & \text{if}\quad E>\Delta
\end{cases},
\end{align}
where we have applied the Andreev approximation. Since the Fermi velocities are the same for all the relevant modes, we can set the normalization factor $N_{e/h,i\alpha}=1$ according to Eq.~\eqref{eqn:S12}. We first consider the upper NS interface and assume an incoming electron state 1:
 \begin{align}
\psi_{N} & =\psi_{N,+}^{e,1}+r_{A1}\psi_{N,+}^{h,1}+r_{A2}\psi_{N,+}^{h,2}+r_{N1}\psi_{N,-}^{e,1}+r_{N2}\psi_{N,-}^{e,2},\\
\psi_{S}^{up} & =c_{1}\psi_{S,+}^{up,1}+c_{2}\psi_{S,-}^{up,1}+c_{3}\psi_{S,+}^{up,2}+c_{4}\psi_{S,-}^{up,2}.
\end{align}
The continuity condition requires $\psi_N(y=W)=\psi_S^{up}(y=W)$. This will lead to the solution:
\begin{equation}
r_{A1}=e^{-i\beta-i\phi/2},\;c_{1}=e^{-i\beta}\;\text{and all other coefficients equal to zero.}
\end{equation}
The same results will be obtained for an incoming electron state 2. For an incoming hole state 1:
\begin{align}
\psi_{N} & =\psi_{N,-}^{h,1}+\tilde{r}_{A1}\psi_{N,-}^{e,1}+\tilde{r}_{A2}\psi_{N,-}^{e,2}+\tilde{r}_{N1}\psi_{N,+}^{h,1}+\tilde{r}_{N2}\psi_{N,+}^{h,2},\\
\psi_{S}^{up} & =c_{1}\psi_{S,+}^{up,1}+c_{2}\psi_{S,-}^{up,1}+c_{3}\psi_{S,+}^{up,2}+c_{4}\psi_{S,-}^{up,2}
\end{align}
Applying the continuity condition, we obtain
\begin{equation}
\tilde{r}_{A1}=e^{-i\beta+i\phi/2},\;c_{2}=e^{i\phi/2}\;\text{and all other coefficients equal to zero.}
\end{equation}
Again, the same result will be obtained for an incoming hole state 2, and we will find that the scattering processes for state 1 and 2 are decoupled.

Now, we turn to the other NS interface and consider an incoming electron state 1:
\begin{align}
\psi_{N} & =\psi_{N,-}^{e,1}+r'_{A1}\psi_{N,-}^{h,1}+r'_{A2}\psi_{N,-}^{h,2}+r'_{N1}\psi_{N,+}^{e,1}+r'_{N2}\psi_{N,+}^{e,2},\\
\psi_{S}^{down} & =c'_{1}\psi_{S,+}^{down,1}+c'_{2}\psi_{S,-}^{down,1}+c'_{3}\psi_{S,+}^{down,2}+c'_{4}\psi_{S,-}^{down,2}
\end{align}
Then we get the solution:
\begin{equation}
r'_{A1}=e^{-i\beta+i\phi/2},\;c'_{2}=e^{-i\beta}\;\text{and all other coefficients equal to zero.}
\end{equation}
For an incoming hole state 1, we have
\begin{align}
\psi_{N} & =\psi_{N,+}^{h,1}+\tilde{r}'_{A1}\psi_{N,+}^{e,1}+\tilde{r}'_{A2}\psi_{N,+}^{e,2}+\tilde{r}'_{N1}\psi_{N,-}^{h,1}+\tilde{r}'_{N2}\psi_{N,-}^{h,2},\\
\psi_{S}^{down} & =c'_{1}\psi_{S,+}^{down,1}+c'_{2}\psi_{S,-}^{down,1}+c'_{3}\psi_{S,+}^{down,2}+c'_{4}\psi_{S,-}^{down,2}
\end{align}
and the solution is
\begin{equation}
\tilde{r}'_{A1}=e^{-i\beta-i\phi/2},\;c'_{1}=e^{-i\phi/2}\;\text{and all other coefficients equal to zero.}
\end{equation}
Then we get following scattering matrix at the NS interfaces:
\begin{equation}
\psi_{out}^{j}=\begin{pmatrix}c_{e,+}^{j}(D)\\
c_{h,-}^{j}(D)\\
c_{e,-}^{j}(U)\\
c_{h,+}^{j}(U)
\end{pmatrix}=\begin{pmatrix}0 & e^{-i\beta-i\phi/2} & 0 & 0\\
e^{-i\beta+i\phi/2} & 0 & 0 & 0\\
0 & 0 & 0 & e^{-i\beta+i\phi/2}\\
0 & 0 & e^{-i\beta-i\phi/2} & 0
\end{pmatrix}\begin{pmatrix}c_{e,-}^{j}(D)\\
c_{h,+}^{j}(D)\\
c_{e,+}^{j}(U)\\
c_{h,-}^{j}(U)
\end{pmatrix}\equiv S_{A}^{j}\psi_{in}^{j}
\end{equation}
where the superscript $j$ denotes the two decoupled modes and $D\,(U)$ denotes the down (up) NS interface. In the normal region, we assume it is transparent, therefore,
\begin{equation}
\psi_{in}^{j}=\begin{pmatrix}c_{e,-}^{j}(D)\\
c_{h,+}^{j}(D)\\
c_{e,+}^{j}(U)\\
c_{h,-}^{j}(U)
\end{pmatrix}=\begin{pmatrix}0 & 0 & e^{-ik_{ej,-}W} & 0\\
0 & 0 & 0 & e^{-ik_{hj,+}W}\\
e^{ik_{ej,+}W} & 0 & 0 & 0\\
0 & e^{ik_{hj,-}W} & 0 & 0
\end{pmatrix}\begin{pmatrix}c_{e,+}^{j}(D)\\
c_{h,-}^{j}(D)\\
c_{e,-}^{j}(U)\\
c_{h,+}^{j}(U)
\end{pmatrix}\equiv S_{N}^{j}\psi_{out}^{j}
\end{equation}
where the scattering matrix describes the phase $\exp{(i\bm{k}\cdot\bm{y})}$ accumulated when a wave packet moves between the SNS junction.

For an Andreev bound state, we have the standing-wave condition $\psi^j_{in}=S_N^jS_A^j\psi^j_{in}$, therefore the bound state energy can be determined by $\text{Det}(\mathbb{I}-S_N^jS_A^j)=0$\cite{Beenakker}. Using the identity
\begin{equation}
\text{Det}\begin{pmatrix}a & b\\
c & d
\end{pmatrix}=\text{Det}(ad-aca^{-1}b),
\end{equation}
we find the equation
\begin{equation}
\text{Det}\left[\mathbb{I}-e^{-2i\beta}T_{+}r_{A}T_{-}r_{A}^{*}\right]=0.
\end{equation}
Here,
\begin{equation}
r_{A}=\begin{pmatrix}0 & e^{-i\phi/2}\\
e^{i\phi/2} & 0
\end{pmatrix},\quad T_{-}=\begin{pmatrix}e^{-ik_{ej,-}W} & 0\\
0 & e^{-ik_{hj,+}W}
\end{pmatrix},\quad T_{+}=\begin{pmatrix}e^{ik_{ej,+}W} & 0\\
0 & e^{ik_{hj,-}W}
\end{pmatrix}.
\end{equation}
Using Eqs.~\eqref{eqn:S13}$-$\eqref{eqn:S16}, after some algebra, we arrive Eq.~\eqref{eqn:subgap spectrum} in the main text:
\begin{equation}
\cos^{-1}\left(\frac{E}{\Delta}\right)=\frac{E\pm E_{Z}}{v_{F}}W\pm\frac{\phi}{2}+n\pi,\quad n\in\mathbb{Z}.
\end{equation}
These equations are related by particle-hole symmetries and the solutions come in pairs as $\pm E$. More specifically,
\begin{align}
\cos^{-1}\left(\frac{E}{\Delta}\right) & =\frac{\pi}{2}\frac{E+E_{Z}}{E_{T}}+\frac{\phi}{2}+n\pi\;\text{and}\;\cos^{-1}\left(\frac{E}{\Delta}\right)=\frac{\pi}{2}\frac{E-E_{Z}}{E_{T}}-\frac{\phi}{2}+n\pi\;\text{give solutions }\pm E_{1},\\
\cos^{-1}\left(\frac{E}{\Delta}\right) & =\frac{\pi}{2}\frac{E-E_{Z}}{E_{T}}+\frac{\phi}{2}+n\pi\;\text{and}\;\cos^{-1}\left(\frac{E}{\Delta}\right)=\frac{\pi}{2}\frac{E+E_{Z}}{E_{T}}-\frac{\phi}{2}+n\pi\;\text{give solutions }\pm E_{2},
\end{align}
where $E_T$ is the corresponding Thouless energy which is given by
\begin{equation}
E_{T}=\frac{\pi}{2}\frac{v_{F}}{W}.
\end{equation}
For the $E=0$ level crossing point, the $\phi-E_{Z}$ relation is:
\begin{equation}
\phi\!\mod2\pi=\pi\left(1\pm\frac{E_{Z}}{E_{T}}\right)\!\mod2\pi.
\end{equation}

\section{setup for the tunneling measurements}

\begin{figure}[!htbp]
\centering
\epsfig{figure=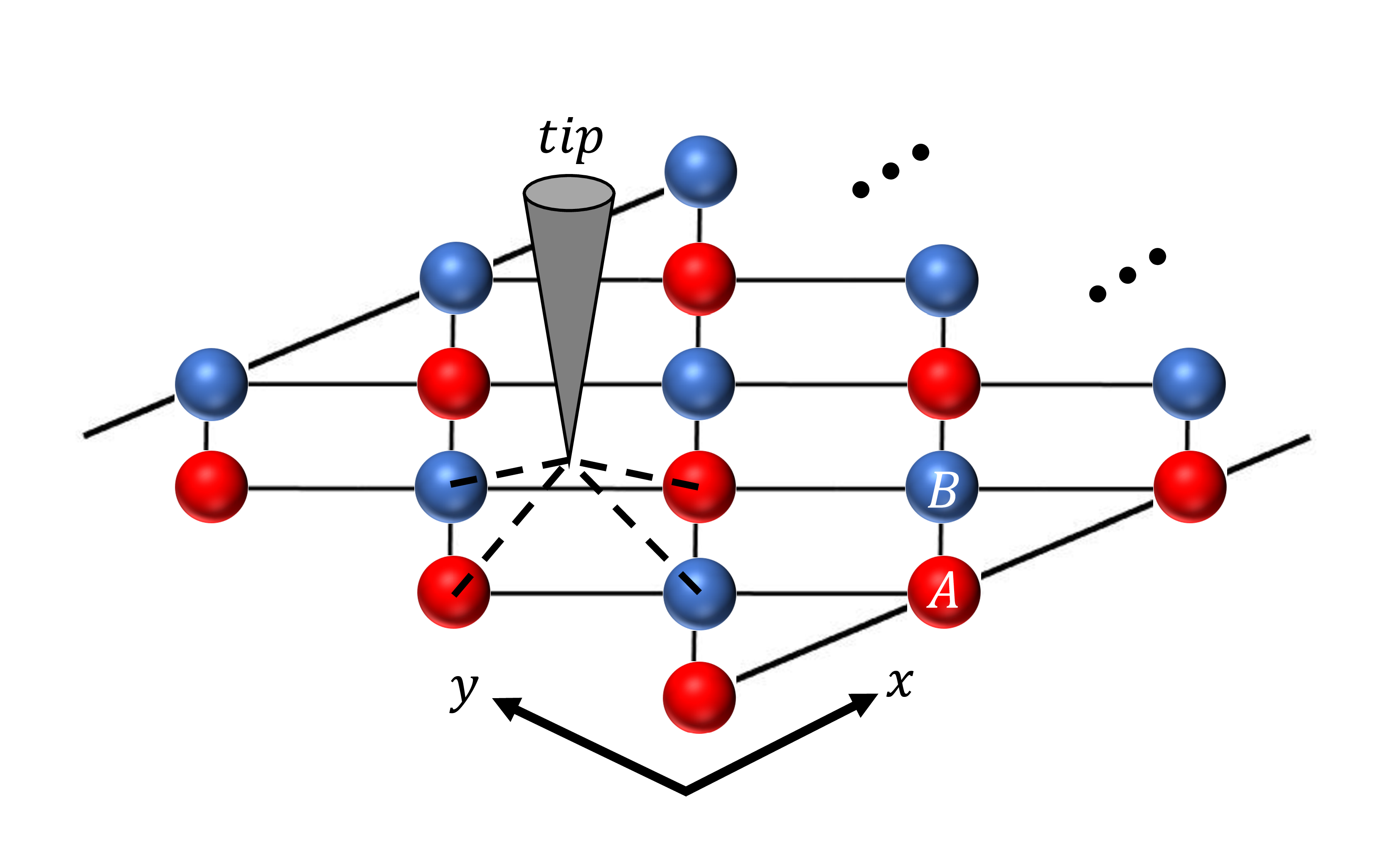, width=0.45\textwidth}	
\caption{The sketch of our STS setup, where the tip is placed above the junction. $A$, $B$ represents the two types of the lattice sites in our model.} \label{fig:STM sketch}
\end{figure}

Fig. \ref{fig:STM sketch} shows the setup for the tunneling measurements on the topological planar JJs.  We assume that the electrons at the apex of the tip can tunnel to the four nearest-neighborhood sites with equal tunneling amplitudes $\tilde{t}$ which can be tuned by adjusting the distance between tip and junction. One can either fix the tip at the ends of the junction to measure the differential conductance arising from the end states or move the tip along the $x$ direction to measure the Fano factor tomography.


To calculate the current and noise\cite{PhysRevB.104.L121406,leeuwen_2013}, we first introduce the Nambu spinor $\psi_{T}=\left(c_{T\uparrow},\,c_{T\downarrow},\,c_{T\downarrow}^{\dagger},-c_{T\uparrow}^{\dagger}\right)^{T}$ and $\psi_{J,i}=\left(c_{J\uparrow i},\,c_{J\downarrow i},\,c_{J\downarrow i}^{\dagger},-c_{J\uparrow i}^{\dagger}\right)^{T}$. We can use Nambu spinors to define Keldysh Green's functions which are $4\times4$ matrices and their components are defined as:
\begin{align}
G_{\alpha,\tau s;\beta,\tau's'}^{R}(t,\,t') & =-i\theta(t-t')\left\langle \left\{ \psi_{\alpha,\tau s}(t),\,\psi_{\beta,\tau's'}^{\dagger}(t')\right\} \right\rangle, \\
G_{\alpha,\tau s;\beta,\tau's'}^{A}(t,\,t') & =-i\theta(t'-t)\left\langle \left\{ \psi_{\alpha,\tau s}(t),\,\psi_{\beta,\tau's'}^{\dagger}(t')\right\} \right\rangle, \\
G_{\alpha,\tau s;\beta,\tau's'}^{<}(t,\,t') & =i\left\langle \psi_{\beta,\tau's'}^{\dagger}(t')\psi_{\alpha,\tau s}(t)\right\rangle, \\
G_{\alpha,\tau s;\beta,\tau's'}^{>}(t,\,t') & =-i\left\langle \psi_{\alpha,\tau s}(t)\psi_{\beta,\tau's'}^{\dagger}(t')\right\rangle,
\end{align}
where $\tau,\tau'$ are particle-hole indices, $s,s'$ are spin indices and $\alpha,\beta$ denote other indices (spatial index $i$ and lead index $T,J$). In the steady state regime, the Green's functions only depend on the time difference $(t-t')$ and can be represented by its Fourier components:
\begin{align}
G(t,\,t') & =\int\frac{\mathrm{d}\omega}{2\pi}e^{-i\omega(t-t')}G(\omega),\\
G(\omega) & =\int\mathrm{d}(t-t')\,G(t,\,t')e^{i\omega(t-t')}.
\end{align}

Using Nambu spinor, the tunneling Hamiltonian~\eqref{tunnelingH} can be written as:
\begin{equation}
H_{\text{tunnel}}=\frac{1}{2}\left[\psi_{T}^{\dagger}\tau_{z}\tilde{t}_{i}\psi_{J,i}\right]+H.c.,
\end{equation}
where we use the Einstein summation convention. In the steady state, the current and noise can be expressed as follows:
\begin{align}
I & =\frac{e\tilde{t}_{i}}{2\hbar}\int\frac{\mathrm{d}\omega}{2\pi}\mathrm{Tr}\left[G_{Ji,T}^{<}(\omega)-G_{T,Ji}^{<}(\omega)\right],\\
S & =\frac{e^{2}\tilde{t}_{i}\tilde{t}_{k}}{\hbar^{2}}\int\frac{\mathrm{d}\omega}{2\pi}\left\{ \mathrm{Tr}\left[G_{T,T}^{<}(\omega)G_{Jk,Ji}^{>}(\omega)-G_{Ji,T}^{<}(\omega)G_{Jk,T}^{>}(\omega)\right]+\mathrm{Tr}\left[G_{Ji,Jk}^{<}(\omega)G_{T,T}^{>}(\omega)-G_{T,Jk}^{<}(\omega)G_{T,Ji}^{>}(\omega)\right]\right\}.
\end{align}

One can use Dyson equations to calculate the Green's functions:
\begin{align}
G_{Ji,Jk}^{R/A} & =g_{Ji,Jk}^{R/A}+g_{Ji,Ji'}^{R/A}\Sigma_{Ji',Jk'}^{R/A}G_{Jk',Jk}^{R/A},\\
G_{T,T}^{R/A} & =g_{T}^{R/A}+g_{T}^{R/A}\Sigma_{T}^{R/A}G_{T,T}^{R/A},\\
G_{Ji,Jk}^{<,>} & =g_{Ji,Jk}^{<,>}+g_{Ji,Ji'}^{R}\Sigma_{Ji',Jk'}^{R}G_{Jk',Jk}^{<,>}+g_{Ji,Ji'}^{R}\Sigma_{Ji',Jk'}^{<,>}G_{Jk',Jk}^{A}+g_{Ji,Ji'}^{<,>}\Sigma_{Ji',Jk'}^{A}G_{Jk',Jk}^{A},\\
G_{T,T}^{<,>} & =g_{T}^{<,>}+g_{T}^{R}\Sigma_{T}^{R}G_{T,T}^{<,>}+g_{T}^{R}\Sigma_{T}^{<,>}G_{T,T}^{A}+g_{T}^{<,>}\Sigma_{T}^{A}G_{T,T}^{A},\\
G_{Ji,T}^{<,>} & =G_{Ji,Jk}^{R}\tilde{t}_{k}\tau_{z}g_{T}^{<,>}+G_{Ji,Jk}^{<,>}\tilde{t}_{k}\tau_{z}g_{T}^{A},\\
G_{T,Jk}^{<,>} & =g_{T}^{R}\tilde{t}_{i}\tau_{z}G_{Ji,Jk}^{<,>}+g_{T}^{<,>}\tilde{t}_{i}\tau_{z}G_{Ji,Jk}^{A},
\end{align}
where the self energies are given by,
\begin{align}
\Sigma_{Ji,Jk}^{R/A} & =\tilde{t}_{i}\tau_{z}g_{T}^{R/A}\tilde{t}_{k}\tau_{z},\\
\Sigma_{Ji,Jk}^{<,>} & =\tilde{t}_{i}\tau_{z}g_{T}^{<,>}\tilde{t}_{k}\tau_{z},\\
\Sigma_{T}^{R/A} & =\tilde{t}_{i}\tau_{z}g_{Ji,Jk}^{R/A}\tilde{t}_{k}\tau_{z},\\
\Sigma_{T}^{<,>} & =\tilde{t}_{i}\tau_{z}g_{Ji,Jk}^{<,>}\tilde{t}_{k}\tau_{z}.
\end{align}
Here the lowercase $g$ represents the isolated Green's functions which are used as the inputs to calculate the set of Dyson equations above. For the metallic tip, we apply the wide band approximation and its isolated Green's functions are given by,
\begin{align}
g_{T}^{R/A} & =\mp i\pi v_{T}\tau_{0}s_{0},\\
g_{T}^{<}(\omega) & =2\pi iv_{T}\left[n_{F}(\omega_{-})\frac{\tau_{0}+\tau_{z}}{2}s_{0}+n_{F}(\omega_{+})\frac{\tau_{0}-\tau_{z}}{2}s_{0}\right],\\
g_{T}^{>}(\omega) & =-2\pi iv_{T}\left[\bar{n}_{F}(\omega_{-})\frac{\tau_{0}+\tau_{z}}{2}s_{0}+\bar{n}_{F}(\omega_{+})\frac{\tau_{0}-\tau_{z}}{2}s_{0}\right],
\end{align}
where $n_F(\omega)$ is the Fermi-Dirac distribution function, $\bar{n}_{F}(\omega)=1-n_F(\omega)$, $\omega_{\pm}=\omega\pm eV$ and $v_T$ is the density of states of the metallic tip at the Fermi level. With the knowledge of $g^R_{Ji,Jk}(\omega)$ in hand, we are able to solve the Dyson equations and obtain the current as well as the noise.

\section{Non-vanishing Zeeman field underneath the SC leads}
In this section, we consider the case that the Zeeman field underneath the SC leads is non-zero. Specifically, we take the extreme case where the Zeeman fields beneath both leads ($E_{Z,L}$) and the junction region ($E_{Z,J}$) are equal. We present the energy spectrum of a $\pi\text{-}0$ junction under open boundary condition for this case. It is noteworthy that the system becomes gapless when $E_{Z,L} > \Delta$, where $\Delta$ is the SC gap.

\begin{figure}[!htbp]
	\centering
	\epsfig{figure=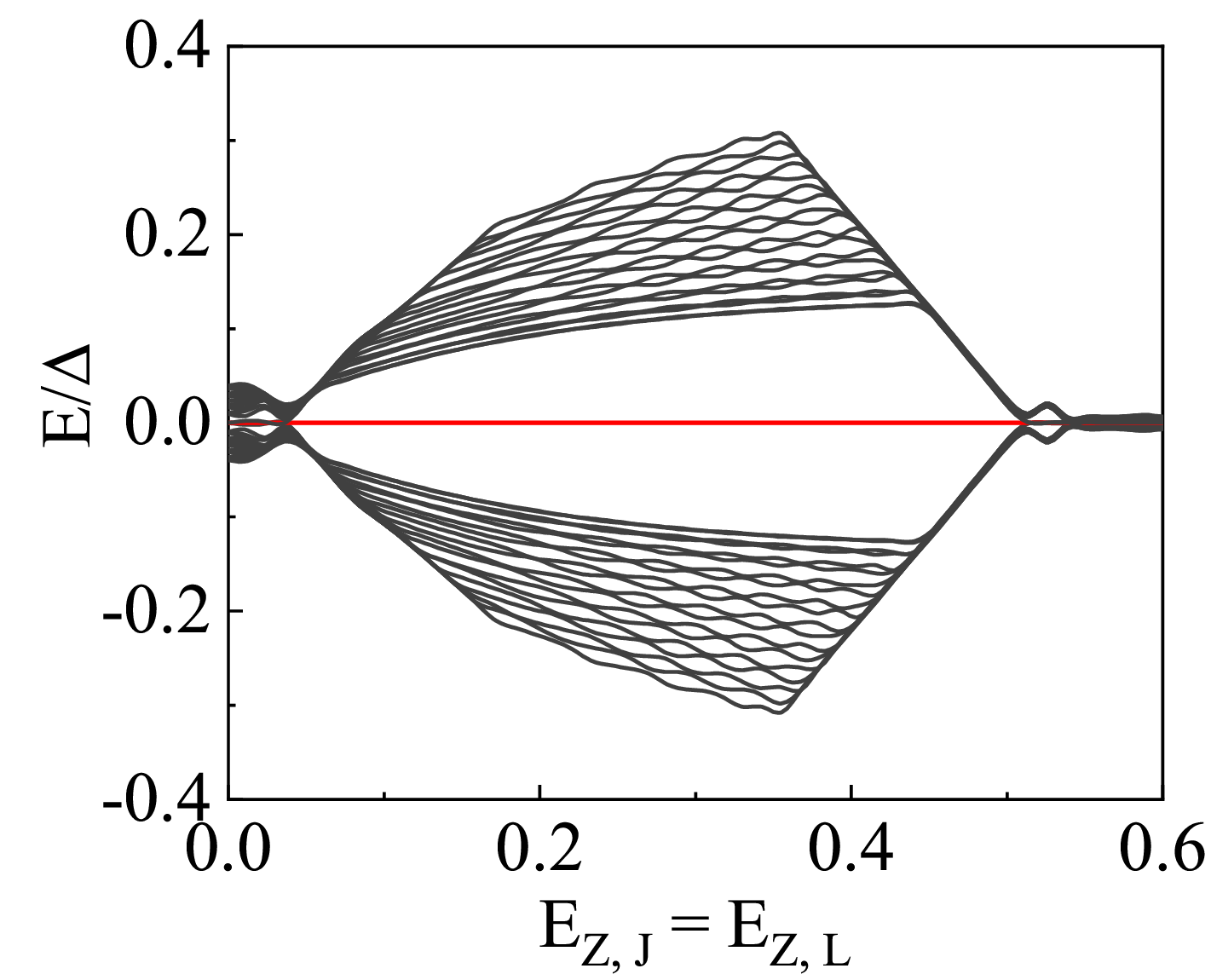, width=0.45\textwidth}
	\caption{Open boundary spectrum of the junction Hamiltonian under the condition that the Zeeman energy in the junction equals to Zeeman energy in the leads (i.e. $E_{Z,J} = E_{Z,L}$). The corresponding parameters are chosen as $\{t, t_1, R, \mu, \phi, \Delta, N_x, N_y, W\} = \{-0.8, 0.4, 0.6, -3, \pi, 0.5, 300, 64, 2\}$}
	\label{fig:EzL=EzJ}
\end{figure}

\end{document}